\documentstyle[prb,multicol,aps,epsf]{revtex}

\begin{document}
\title{Impurity-induced polaritons in a one-dimensional chain\footnote{ J. Opt. Soc. Amer. B, 2000, 17, No. 9, 1498-1508.}}
\author{A. Yamilov,$^{\dagger}$ L.I. Deych$^{\ddagger}$ and A.A. Lisyansky$^{\dagger}
$}
\address{$^{\dagger}$Department of Physics, Queens College of CUNY, Flushing, NY 11367}
\address{$^{\ddagger}$Department of Physics, Seton Hall University, 400 South Orange
Ave, South \\
Orange, NJ 07079}
\date{\today}
\maketitle

\begin{abstract}
A detailed analytical study of an impurity induced polariton band arising
inside a spectral gap between lower and upper polariton branches is
presented. Using the microcanonical method, we calculate the density of
states and localization length of the impurity polaritons. Analytical
results are compared with numerical simulations and excellent agreement is
found.
\end{abstract}

\begin{multicols}{2}
\section{Introduction}

An opportunity to create Anderson localized states of electromagnetic
excitations in disordered dielectrics has attracted a great deal of
attention during the last two decades.\cite{John,Ad} Though the localization of light in
three-dimensional uniform-on-average random systems proved to be difficult
to achieve, new opportunities opened up with development of the structures
with periodically modulated dielectric properties, so called photonic
crystals.\cite{Yabl-87,Photonics-Book} It was first suggested in Ref. %
\onlinecite{S.John-87} that the photon localization can be more easily
achieved if the photon density of states (DOS) is depleted at certain
frequency domains. Photonic crystals allow the creation of photon band-gaps,
regions of frequencies where the photon DOS is zero. Introducing isolated
defects in such structures, one can create local photon modes similar to
well known defect phonon modes in regular crystals\cite
{Lifshitz,LifshitzKosevich,Lifshitz Kirpichnikov,Maradudin,MandM} or
electron impurity states in semiconductors.\cite{Efros} A forbidden gap in
the photonic spectrum occurs when the electromagnetic wave length becomes
comparable to the lattice constant. Consequently, for microwave, infrared
and visible ranges, the photonic crystal is a structure of macroscopic
dimensions. Local states in these structures also arise due to defects of
macroscopical dimensions. The great interest in photonic crystals is
promoted by their highly unusual quantum electrodynamic properties. For
instance, spontaneous emission, which is completely suppressed within the
frequency range of the band gaps,\cite{sontemiss} can be effectively
controlled by introducing local modes.\cite{PhC,Joannopoulos}

The intriguing possibilities of photonic crystals initiated interest in
optical effects in other types of photonic band-gaps. Such gaps can arise,
for example, between different polariton branches. The fact that the
frequency region between polariton branches can represent a stop-band for
electromagnetic waves propagating in certain directions has been well known
for a long time. The reflection coefficient in this situation can reach the
magnitude of 90 percent or even greater. This effect was used in the 40's
and 50's in order to create monochromatic infrared beams (it was called the
method of residual or reststrahlen rays, and the respective spectral
interval was called the reststrahlen region). In certain cases, however, the
stop bands exist for all the directions, and then a genuine spectral gap
arises with the same consequences for quantum electrodynamics as in the case
of photonic crystals. This was first realized in Ref. \onlinecite{Rupasov},
where it was shown that a two-level atom can form an atom-field coupled
states with suppressed emission similar to the states discussed in Ref. %
\onlinecite{dressed-atom} for photonic crystals. Periodic arrangement of
such two-level systems produces an impurity-induced polariton band within
the polariton gap.\cite{Singh}

Local polariton states associated with a regular isotopic defect without any
intrinsic optical activity were introduced in Refs. %
\onlinecite{Classics,Podolsky}. These states are coupled states of
electromagnetic excitations with phonons (or excitons), with both
components, including the electromagnetic component, being localized in the
vicinity of the defect. These states are, in a certain sense, analogous to
local photons in photonic crystals considered in Refs. 
\onlinecite
{PhC,Joannopoulos}, although they arise due to microscopic impurities in
regular crystal lattices. On the other hand, local polaritons can be
considered as local excitations of a crystal coupled with the
electromagnetic field. Regular local phonons (excitons) also interact with
the electromagnetic field, but this interaction results mainly in absorption
of light and radiative decay of the states.\cite{MandM} The local polaritons
arise in the region where electromagnetic waves do not propagate, and there
are, therefore, neither defect induced absorption of light nor radiative
damping of the local states. The electromagnetic interaction leads in this
case to new optical effects, and strongly affects the properties of the
local states. It is well known, for instance, that local phonons or excitons
in a 3D system (as well as local photons in photonic crystals) arise only if
the ``strength'' of a defect exceeds a certain threshold. Local polaritons
in systems with an isotropic dispersion split off the allowed band without a
threshold even in three dimensions.\cite{Classics} This effect is caused by
the interaction with electromagnetic field, which results in the Van-Hove
singularity in the polariton DOS.

One of the optical effects caused by local polaritons is resonant tunneling
of electromagnetic wave with gap frequencies. This effect was first
suggested in Ref. \onlinecite{Deych}, where the results of numerical
simulation of electromagnetic wave propagation through a relatively short 1D
chain with a single defect were reported. It was found that the transmission
coefficient at the frequency of the local polariton state increases
dramatically up to the a value close to unity. In spite of the general
understanding that local states should produce local tunneling, this result
still seems surprising because transmission of light is effected by a defect
with microscopic dimensions much smaller than the light's wavelength. In
addition, the energy of the electromagnetic component of local polaritons is
much smaller than that of the phonon component. Traditional wisdom based
upon properties of conventional propagating polaritons tells that, in such a
situation, most of the incident radiation must be simply reflected. The
results of Ref. \onlinecite{Deych} demonstrated that this logic does not
apply to local polaritons. The physical explanation of the result can be
given on the basis of consideration of local polaritons as the result of
interaction between electromagnetic field and local phonons. The latter have
macroscopic dimensions comparable with light's wavelength, making the
interaction effective. An electromagnetic wave is carried through a sample
by phonons, which is tunneling resonantly due to the local state with mostly
phonon contribution. Tunneling was confirmed in Refs. %
\onlinecite{OurEuroPhysLett,Blah}, where we analytically calculated the
transmission coefficient through a linear chain with a single defect. We
found that a light impurity indeed gives rise to the resonance transmission
through the forbidden gap of the polariton spectrum. The resonance occurs
when the defect is placed at the center of the chain. The transmission at
the resonance becomes independent of the chain's length and reaches the
value of unity if the resonance frequency coincides with the center of the
polariton gap. The analytical calculations provided an explanation for the
numerical results of Ref. \onlinecite{Deych}, in particular it explained a
strongly asymmetrical frequency profile of the transmission coefficient, and
the presence of a deep minimum following directly after the maximum.

Analytical calculations in Ref. \onlinecite{Blah} were supplemented with
numerical simulations of the electromagnetic wave transmission through the
chain with a finite concentration of impurities. The simulations
demonstrated the development of a defect-induced polariton band within the
forbidden polariton gap. In this paper, we present results of an analytical
study of properties of this band. Neglecting spatial dispersion of phonons,
we are able to obtain analytical expressions for DOS and the Lyapunov
exponent for the band in the limit of large concentration of impurities $%
l_{0}/l_{def}\gg 1$, where $l_{def}$ is the average distance between
impurities, and $l_{0}$ is the localization length of the single local
polariton state. In this limit the method of microcanonical ensemble,
complemented by expansion over $\left( l_{0}/l_{def}\right) ^{-1},$ gives
the DOS of an effective system with a uniform distribution of impurities,
and describes its renormalization caused by local fluctuations of impurity
positions. It also reveals localization of the states of the band, which are
all localized due to the one-dimensional nature of the model. The properties
of this polariton band are found to be drastically different from that of
the band produced by two-level atoms.\cite{Singh} The analytical results
show excellent agreement with numerical calculations and lead to suggestions
regarding experimental observation of the impurity induced polariton bands.

\section{The Model of One-dimensional Polaritons And The Method of
Calculations}

\subsubsection{The model}

The main objective of the present paper is to study properties of the
impurity induced polariton band arising within the polariton gap of a one
dimensional chain of atoms with dipole moments $P_{n}$, where $n$ refers to
a position of an atom in the chain. Dynamics of the chain interacting with a
scalar ``electromagnetic field'' $E(x)$ is described within the
tight-binding approximation with an interaction between nearest neighbors
only, 
\begin{equation}
(\Omega _{n}^{2}-\omega ^{2})P_{n}+\Phi (P_{n+1}+P_{n-1})=\alpha E(x_{n}),
\label{dipoles}
\end{equation}
where $\Phi $ is the parameter of the interaction between the dipoles, $%
\alpha $ is a coupling parameter between the dipoles and the electromagnetic
field,\ and $\Omega _{n}^{2}$ represents a site energy. We shall assume that
the chain is composed of atoms of two different types, which differ in their
site energy only. In this case $\Omega _{n}^{2}$ can be presented in the
following form: 
\begin{equation}
\Omega _{n}^{2}=\Omega _{0}^{2}+\Delta \Omega ^{2}c_{n},  \label{site_energy}
\end{equation}
where $\Delta \Omega ^{2}=\Omega _{1}^{2}-\Omega _{0}^{2}$, $\Omega _{0}^{2}$
and $\Omega _{1}^{2}$ are site energies of the respective atoms, and $c_{n}$
is a random variable which takes values 0 and 1 with probabilities $1-p$ and 
$p, $ respectively. We assume that $p\ll 1$ so that $\Omega _{0}$ is
attributed to host atoms, whereas $\Omega _{1}$ corresponds to the
impurities. Eq. (\ref{dipoles}) can be interpreted in terms of both
exciton-like and phonon-like excitations. In the latter case, $\Delta \Omega
^{2}=(1-M_{imp}/M_{host})\omega ^{2}$, where $M_{host}$ and $M_{imp}$ are
the masses of host atoms and impurities, respectively.

Polaritons in the system arise as collective excitations of dipoles
(polarization waves) coupled to the scalar electromagnetic wave, $E(x_{n})$.
The equation of motion of the electromagnetic component is given by the
Maxwell equation: 
\begin{equation}
\frac{\omega ^{2}}{c^{2}}E(x)+\frac{d^{2}E}{dx^{2}}=-4\pi \frac{\omega ^{2}}{%
c^{2}}\sum_{n}P_{n}\delta (na-x),  \label{Maxwell}
\end{equation}
where $c$ is the speed of the electromagnetic wave in vacuum and $a$ is the
distance between two nearest-neighbor dipoles. Eqs. (\ref{dipoles}) and (\ref
{Maxwell}) present a {\it microscopic} description of the transverse
electromagnetic waves propagating along the chain in the sense that it does
not make use of the concept of the dielectric permeability, and takes into
account all modes of the field including those with wave numbers outside of
the first Brillouin band.

Considering propagation of the electromagnetic wave as a sequence of
scattering events separated by free propagation, one can present Eq. (\ref
{Maxwell}) in a discrete form. The entire system of dynamic equations (\ref
{dipoles}) and (\ref{Maxwell}) can then be transformed into a
transfer-matrix form: 
\begin{equation}
v_{n+1}=\hat{\tau}_{n}v_{n},  \label{EP}
\end{equation}
Vectors $v_{n}$ with components $(P_{n},P_{n+1},E_{n},E_{n}^{\prime }/{k)}$,
where $E_{n}$ and $E_{n}^{\prime }$ are electromagnetic field and its
derivative at the $n$-th site, represent the state of the system. In
general, the matrix $\hat{\tau}_{n}$, which describes the field propagation
between adjacent sites, is a $4\times 4$ matrix.\cite
{Deych,OurEuroPhysLett,Blah} This makes an analytical consideration of Eq. (%
\ref{EP}) very cumbersome, even in the case of a single impurity. At the
same time, numerical simulations carried out in Refs. %
\onlinecite{OurEuroPhysLett,Blah} showed that most of the features of the
impurity induced polariton band are well reproduced even when the
interaction between the dipoles (spatial dispersion) is neglected. In this
approximation, the two first components of the dynamic vector $v_{n}$ are
easily expressed in terms of the other two, and the transfer matrix $\hat{%
\tau}_{n}$ becomes a $2\times 2$ matrix of the following form: 
\begin{equation}
\hat{\tau}_{n}=\left( 
\begin{array}{cc}
\cos ka & \sin ka \\ 
-\sin ka+\beta _{n}(ka)\cos ka & \cos ka+\beta _{n}(ka)\sin ka
\end{array}
\right) .  \label{T reduced}
\end{equation}
The parameter $\beta _{n}$

\begin{equation}
\beta _{n}=\frac{d^{2}}{\omega ^{2}-\Omega _{n}^{2}};  \label{polarizability}
\end{equation}
with $d^{2}=4\pi \alpha /a$ represents the polarizability of the $n$-th
dipole.

The main object of interest in studies of one-dimensional disordered systems
is the Lyapunov exponent (LE), $\lambda $, which in our case can be defined
as\cite{MicrocanMethod} 
\begin{equation}
\lambda =\lim_{{_{L\rightarrow \infty }}}\frac{1}{L}\ln \frac{\left\|
\prod_{1}^{N}\hat{\tau}_{n}v_{0}\right\| }{\left\| v_{0}\right\| },
\label{LE}
\end{equation}
where $L$ is the total length of the chain consisting of $N$ atoms. It is
well-known that the quantity defined by Eq. (\ref{LE}) is non-random
(self-averages). It characterizes the spatial extent of the envelope of
system eigenstates, all of which are localized in one dimension (see Ref. %
\onlinecite{LGP} and references therein). The same quantity also describes
the typical value of the transmission coefficient, $T,$ of an external
excitation incident upon the system: $T_{typ}\simeq \exp (-\lambda L)$, and
through the Thouless relation\cite{Thouless} it determines the DOS of the
system. Calculation of LE in the spectral region of the polariton gap of the
pure system is the main task of this section.

\subsubsection{The Method of Microcanonical Ensemble}

The major problem with calculating of LE or DOS in the region of impurity
bands is that a simple concentration expansion is not able to describe the
impurity band. Electron and phonon impurity bands have been intensively
studied in the past (see, for example, Ref. \onlinecite{LGP}). In the limit
of extremely small concentration of impurities, $\left( pl_{0}\right) /a\ll
1,$ it was proved possible to provide a regular systematic of the arising
states and use statistical arguments to describe their DOS.\cite{Lifshitz}
Another method used for this type of calculations employed the so-called
phase formalism (see Ref. \onlinecite{LGP} and references therein). This
approach allowed one to calculate DOS of impurity bands in different
spectral regions, including exponential behavior at the tails of the band
for both small and large, $(pl_{0})/a\gg 1,$ concentrations. In the case of
local polaritons the localization length, $l_{0}$, of an individual local
state can be so large that one has to deal with the situation described by
the latter inequality even in the case of a rather small concentration of
impurities. Therefore, in this paper we focus mainly upon the properties of
the well-developed polariton impurity band when individual local states are
overlapped. To this end we use the method of the microcanonical ensemble,
which was first suggested for analytical calculations of LE in a
one-dimensional single-band model of a disordered alloy.\cite{MicrocanMethod}
The advantage of this method over other methods of DOS calculations is
that it allows one to calculate simultaneously both DOS and the localization
length. Its main shortcoming is that, as we shall see later, this method is
a version of ``effective media'' methods, and as such it is unable to
describe DOS near fluctuation boundaries of the spectrum. At the same time
the results obtained allow a clear physical interpretation and, as follows
from comparison with numerical simulations, they quite accurately describe
properties of the bulk of the impurity band.

The starting point of our calculations is the general definition of LE given
by Eq. (\ref{LE}). In spite of LE being a self-averaging quantity, it is
convenient in practical calculations to perform averaging over the random
configurations of the impurities. The regular ensemble of the realization is
described by the fixed concentration of the impurities $p$, while their
total number can vary from realization to realization. Therefore one can
distinguish two causes for fluctuations: (1) local arrangements of the
impurities and (2) the total number of the impurities. The main idea of the
method of microcanonical ensemble is to reduce the finite size fluctuations
in the system, eliminating the fluctuations of the total number of
impurities. Such an ensemble with a fixed number of impurities is called the
microcanonical ensemble in analogy with statistical physics. At the same
time the result of the averaging in the limit $L\rightarrow \infty $ should
not depend upon the type of ensemble used by virtue of the self-averaging
nature of LE. The key idea of the microcanonical method is based on the
assumption that with one cause of fluctuations eliminated, one can obtain
reliable results when the microcanonical ensemble average of $\langle \ln
(...)\rangle $ is replaced by $\ln \langle ...\rangle $. Such a substitution
gives an exact result in the case of commutating matrices, and leads to an
excellent agreement between analytical calculations and simulations in the
case of $2\times 2$ matrices with a single band spectrum.\cite{MicrocanMethod}

Using the microcanonical ensemble, one can evaluate the average over all
matrix sequences using the following expression derived in Ref. 
\onlinecite
{MicrocanMethod}: 
\begin{equation}
\left\langle \prod_{1}^{N}\hat{\tau}_{n}\right\rangle ={{{%
{N \choose pN}%
}}}^{-1}\frac{1}{\left( pN\right) !} \left. \frac{\partial ^{pN}}{%
\partial x^{pN}}\left( \hat{\tau}_{0}+x\hat{\tau}_{1}\right) ^{N}\right|
_{x=0},  \label{binomial}
\end{equation}
where $\hat{\tau}_{0}$ and $\hat{\tau}_{1}$ are host and defect matrices,
respectively and $x$ is a free parameter. The derivative in Eq. (\ref
{binomial}) then can be presented in the form of the Cauchy integral 
\begin{equation}
\left\langle \prod_{1}^{N}\hat{\tau}_{n}\right\rangle ={{%
{N \choose pN}%
}}^{-1}\frac{1}{2\pi i}\int_{C}dx\frac{\nu ^{N}\left( x\right) }{x^{pN+1}}%
\hat{D}\left( x\right) ,  \label{Cauchy}
\end{equation}
where the contour of integration $C$ is taken on the complex plane around $%
x=0$, $\nu \left( x\right)$ is the largest eigenvalue of the matrix $%
\hat{\tau}_{0}+x\hat{\tau}_{1}$, and the matrix $\hat{D}( x) ,$   
\begin{equation}
\hat{D}(x)=\nu ^{-N}( x) \left( \hat{\tau}_{0}+x\hat{\tau}%
_{1}\right) ^{N},  \label{pol}
\end{equation}
has eigenvalues not exceeding $1$ in absolute value. In the limit of large $N
$ the above integral can be evaluated by the saddle point method. As the
result we arrive at the following expression for the complex valued LE, $%
\widetilde{\lambda }$: 
\begin{eqnarray}
\widetilde{\lambda } &=&\lim_{{_{N\rightarrow \infty }}}\frac{1}{L}\ln \left[
\text{largest eigenvalue }\left\langle \prod_{1}^{N}\hat{\tau}%
_{n}\right\rangle \right] \nonumber \\
&=&\frac{1}{a}\left[ \ln \nu ( x_0) -( 1-p) \ln \frac {x_0}{1-p} 
 + p\ln p \right], \label{SaddlePoint}
\end{eqnarray}
where $x_{0}$ is defined by the saddle point equation 
\begin{equation}
\left. \frac{\partial \ln \nu \left( x\right) }{\partial \ln x}\right|
_{x=x_{0}}=p.  \label{x0}
\end{equation}
The real part of $\widetilde{\lambda }$ given by Eq. (\ref{SaddlePoint})
represents LE, while its imaginary part according to Thouless\cite{Thouless}
gives the integral density of states $N(\omega )$ in the impure system: 
\begin{eqnarray}
\lambda  &=&\mathop{\rm Re}\left[ \widetilde{\lambda }\right] ,
\label{Re Lyamda} \\
N(\omega ) &=&-\frac{1}{\pi }\mathop{\rm Im}\left[ \widetilde{\lambda }%
\right] .  \label{Im Lyamda}
\end{eqnarray}

The eigenvalues of the matrix $\hat{\tau}_{0}+x\hat{\tau}_{1}$ can be found
from the equation 
\begin{equation}
\frac{\nu }{1+x}+\frac{1+x}{\nu }=\kappa \left( x\right) =2\cos \left(
ka\right) +\frac{\beta _{0}+x\beta _{1}}{1+x},  \label{Eigenvalues}
\end{equation}
where $\beta _{0}$ and $\beta _{1}$ are polarizabilities of the host and
defect atoms, respectively.

It is convenient to rewrite Eqs. (\ref{SaddlePoint}) and (\ref{Eigenvalues})
in terms of a new variable $y=x/(1+x)$: 
\begin{eqnarray}
\kappa (y) &=&(1-y)\nu (y)+\frac{1}{(1-y)\nu (y)}=\kappa _{0}(\omega
)+yF(\omega ),  \label{InTermsOfY} \\
\kappa _{0}(\omega ) &=&2\cos \left( ka\right) +d^{2}ka\sin (ka)\frac{1}{%
\omega ^{2}-\Omega _{0}^{2}},  \label{kappa} \\
F(\omega ) &=&d^{2}ka\sin (ka)\frac{\Omega _{1}^{2}-\Omega _{0}^{2}}{\left(
\omega ^{2}-\Omega _{0}^{2}\right) \left( \omega ^{2}-\Omega _{1}^{2}\right) 
}.  \label{F}
\end{eqnarray}
In the long-wave limit $\epsilon =ka\ll 1$ functions $F(\omega )$ and $%
\kappa _{0}(\omega )$ can be simplified and presented in the form, which
clarifies their physical meaning: 
\begin{eqnarray*}
F(\omega ) &\simeq &\epsilon ^{2}f(\omega ), \\
\kappa _{0}(\omega ) &\simeq &2+\epsilon ^{2}\gamma (\omega ),
\end{eqnarray*}
where $f(\omega )$ is defined according to 
\begin{equation}
f(\omega )\simeq \frac{d^{2}(\Omega _{1}^{2}-\Omega _{0}^{2})}{(\omega
^{2}-\Omega _{0}^{2})(\omega ^{2}-\Omega _{1}^{2})}  
\label{f(omega)}
\end{equation}
and represents the difference $\beta _{1}-\beta _{0}$ between
polarizabilities (\ref{polarizability}) of the impurities and host atoms.
Function $\gamma (\omega )$ defined as 
\begin{equation}
\gamma (\omega )=1+\frac{d^{2}}{\Omega _{0}^{2}-\omega ^{2}}  \label{gamma}
\end{equation}
is the longwave dielectric function of the pure chain.

Similarly, Eq. (\ref{SaddlePoint}) transforms into 
\begin{eqnarray}
\widetilde{\lambda }(p) &=& \ln \left[ (1-y_{0}(p))\nu \left( y_{0}(p)\right) \right]  
\nonumber  \\  
&- & \left[ p\ln \frac{y_{0}\left( p\right) }{p}+\left( 1-p\right) \ln 
\frac{1-y_{0}\left( p\right) }{1-p}\right] ,  
\label{Lyapunov}
\end{eqnarray}
with the saddle point equation reading as 
\begin{equation}
\pm \left. \frac{y(1-y)F}{\sqrt{\left( \kappa _{0}+yF\right) ^{2}-4}}%
+y\right| _{y=y_{0}}=p,  \label{y}
\end{equation}
The choice of the sign in Eq. (\ref{y}) is determined by the requirement to
use the greatest of the eigenvalues $\nu $. Introduction of the new variable 
$y$ turns $\kappa \left( y\right) $ in a linear function of $y$ essentially
simplifying future calculations.

\section{The Lyapunov exponent and the density of states}

\subsubsection{Boundaries of the impurity band}

Dielectric function $\gamma (\omega )$ determines the frequency region of
the polariton gap: for $\Omega _{0}^{2}<\omega ^{2}<\Omega _{0}^{2}+d^{2}$
it is negative, and hence, propagating modes do not exist in this region. We
shall assume that the defect frequency $\Omega _{1}^{2}$ obeys the
inequality $\Omega _{0}^{2}<\Omega _{1}^{2}<\Omega _{0}^{2}+d^{2}$, so that
the impurity induced band develops inside the gap of the original spectrum.
As we already mentioned, the approach we use in the paper belongs to the
class of effective-medium approximations since we neglect here certain kinds
of fluctuations in the system. It is natural to expect, therefore, that
within this approach the impurity band would have well-defined spectral
boundaries outside of which the differential DOS remains exact zero. Our
first goal is to determine these boundaries and find the concentration
dependence of the width of the impurity band. The differential DOS, $\rho
(\omega )=dN(\omega )/d\omega ,$ takes on non-zero values when $\widetilde{%
\lambda }(\omega )$ acquires a non-constant imaginary part. Rewriting Eq. (%
\ref{Lyapunov}) in the form 
\begin{eqnarray}
\widetilde{\lambda }(\omega ) &= & \frac{1}{a} \ln \left[ \frac{1}{2}
\left( \kappa _{0}+y_{0}F\pm \frac{y_{0}(1-y_{0})F}{p-y_{0}}\right) \right]  \nonumber \\
&- & \frac{1}{a} \left[ p\ln \frac{y_{0}}{p}+\left( 1-p\right) \ln \frac{1-y_{0}}{1-p}\right],  
\label{Lyap Compl}
\end{eqnarray}
one can see that in the case of real $y_{0}$, $\mathop{\rm Im}\widetilde{%
\lambda }(\omega )$ can be either zero or $\pi $ (the latter happens when
the argument of $\ln $ is negative). In both cases differential DOS is
obviously zero, and it can only take on a non-zero value when $y_{0}$
becomes complex. Eq. (\ref{y}), which defines $y_{0}$, is formed by a
polynomial of the third order with real coefficients; therefore it has
either three real roots or one real solution and a complex conjugated pair.
In order to describe formation of the polariton band, one has to select the
root, which has a complex component in a certain frequency interval and
yields positive LE. Comparison with numerical simulations shows that the
choice of only one of three saddle point, according to above mentioned
criteria, produces the correct description of the impurity band. It is not
difficult to show that at the frequency where $y_{0}(\omega )$ becomes
complex, the derivative $\partial y_{0}/\partial \omega $ diverges. This
fact gives us an explicit equation for the spectrum boundaries. From Eq. (%
\ref{y}) one can find that the divergence occurs when $y_{0}$ satisfies the
equation 
\begin{eqnarray}
y_{0}^{3}F\left[ F(1-p)+\kappa _{0}\right] +y_{0}^{2}\left[ \kappa
_{0}F(1-3p)+k_{0}^{2}-4\right] 
\nonumber \\
+y_{0}\kappa _{0}\left( F-2\kappa
_{0}^{2}+4\right) +p(\kappa _{0}^{2}-4)=0.  
\label{y on a boundary}
\end{eqnarray}
Eqs. (\ref{y}) and (\ref{y on a boundary}) define the concentration
dependence of the boundaries of the polariton impurity band. An approximate
solution of this equation can be obtained as a formal series in powers of
parameter $\epsilon =ka$. It will be seen, however, from the results that
the actual expansion parameter in this case is $1/pl_{0}\ll 1$. The solution
of Eq. (\ref{y}) with the accuracy to $\epsilon ^{2}$ can be obtained in the
form: 
\begin{eqnarray}
y=p&-&p(1-p)\frac{f}{2\sqrt{pf-\gamma }}\epsilon 
\label{y expansion} \\
&+&p(1-p)\frac{2\left(
pf-\gamma \right) f\left( 1-2p\right) -fp\left( 1-p\right) }{8\left(
pf-\gamma \right) ^{2}}f^{2}\epsilon ^{2},  
\nonumber
\end{eqnarray}
where $y=p$ is the only non-vanishing zero order approximation for $y$.
Since two other solutions, $y=0,$ correspond to a singular point of the
integral (\ref{Cauchy}), the chosen solution $y=p$ represents the only
saddle point accessible within our perturbation scheme. Using additional
criteria outlined above we verify that the solution given by Eq. (\ref{y
expansion}) correctly reproduces the behavior of LE. Substituting Eq. (\ref
{y expansion}) into Eq. (\ref{Lyapunov}) one finds the complex LE in the
long wavelength approximation: 
\begin{equation}
\widetilde{\lambda }(\omega )=\frac{\omega }{c}\left( \sqrt{pf(\omega
)-\gamma (\omega )}-\frac{p(1-p)f^{2}(\omega )}{8[pf(\omega )-\gamma (\omega
)]}\epsilon \right) .  \label{FinalLyapun}
\end{equation}
\bigskip According to Eq. (\ref{FinalLyapun}), $\widetilde{\lambda }(\omega
) $ acquires an imaginary part at frequencies obeying the inequality 
\begin{equation}
pf(\omega )-\gamma (\omega )\leq 0,  \label{tentative boundaries}
\end{equation}
The boundaries are determined by the respective equation $pf(\omega )-\gamma
(\omega )=0$, which coincides with the long wavelength limit of Eq. (\ref{y
on a boundary}). {This equation determines two points where }$pf(\omega
)-\gamma (\omega )$ changes sign to negative{: 
\begin{eqnarray}
\omega _{il,pu}^{2} &=&\frac{1}{2}(\Omega _{0}^{2}+\Omega _{1}^{2}+d^{2}) 
\label{Omega24} \\
&\pm  & \frac{1}{2}\sqrt{(\Omega _{0}^{2}+d^{2}-\Omega _{1}^{2})^{2}+4d^{2}(\Omega
_{1}^{2}-\Omega _{0}^{2})p},
\nonumber  
\end{eqnarray}
The first of these solutions belongs to the initial polariton band-gap and
as such represents the low-frequency boundary of the new impurity band. The
second one lies outside of the gap and is a bottom frequency, modified by
impurities, of the upper polariton branch of the initial spectrum. These two
frequencies, however, are not the only points where the expression }$%
pf(\omega )-\gamma (\omega )$ turns negative. Two other points are 
\begin{equation}
\omega _{pl}^{2}=\Omega _{0}^{2},\ \ \ \ \  \omega _{iu}^{2}=\Omega _{1}^{2},  \label{Omega13} \\
\end{equation}
{\ and the change of the sign at these points occurs through infinity of \ }$%
pf(\omega )-\gamma (\omega )$ rather than through zero. These two
frequencies do not depend upon the concentration of impurities and present,
therefore, stable genuine boundaries of the spectrum. This property of {$%
\Omega _{0}$ and $\Omega _{1}$} is due to their resonance nature (they
correspond to the poles of the respective polarizabilities) and disappears
when, for example, the spatial dispersion is taken into account. At the same
time, the numerical simulations of Ref. \onlinecite{Blah}, in which the
spatial dispersion was taken into consideration, indicate that the shift of
these frequency from their initial values is negligibly small for realistic
values of the inter-atom interaction parameter $\Phi ,$ even for a
relatively large concentration of impurities.

These four frequencies set the modified boundaries of the initial
polariton spectrum of the pure system ($\omega _{pl}^{2}$, $\omega
_{pu}^{2})$, and boundaries of the newly formed impurity band ($\omega
_{iu}^{2}$, $\omega _{il}^{2}$). The lower boundary of the forbidden gap of
the crystal, $\Omega _{0}$, is not affected by the impurities - the
singularity in the polariton DOS at this point survives for any
concentrations of defects. The upper boundary of the band gap, $\omega _{pu}$%
, shifts toward higher frequencies with the concentration and when $p=1$, it
coincides with the upper band boundary of the new crystal, $\Omega _{L}^{i}=%
\sqrt{\Omega _{1}^{2}+d^{2}}.$ Frequencies $\omega _{il}$ and $\omega _{iu}$
give approximate values for the lower and upper boundaries of the impurity
induced pass band which arises inside the original forbidden gap $\Omega
_{0}<\omega _{il}<\omega _{iu}<\sqrt{\Omega _{0}^{2}+d^{2}}$. The impurity
band grows asymmetrically with concentration: while its lower boundary moves
towards $\Omega _{0}$ with an increase of the concentration, the upper edge
remains fixed at $\omega _{iu}=\Omega _{1}$. Such a behavior of the
impurity band agrees well with our numerical results.\cite{Blah}

The width of the impurity induced band defined in terms of squared
frequencies $\Delta _{im}^{2}=${$\omega _{iu}^{2}-\omega _{il}^{2}$ }can be
found from Eqs. (\ref{Omega24}) and (\ref{Omega13}) as 
\begin{eqnarray}
\Delta _{im}^{2}&=&\frac{1}{2}\sqrt{(\Omega _{0}^{2}+d^{2}-\Omega
_{1}^{2})^{2}+4d^{2}(\Omega _{1}^{2}-\Omega _{0}^{2})p}
\nonumber \\
&-&\frac{1}{2}(\Omega
_{0}^{2}+d^{2}-\Omega _{1}^{2}).  \label{width}
\end{eqnarray}
For $\Omega _{1}$ not very close to the upper boundary of the initial
polariton gap, $\Omega _{L}=\sqrt{\Omega _{0}^{2}+d^{2}},$ the linear in $p$
approximation is sufficient to describe the concentration dependence of the
band width 
\begin{equation}
\Delta _{im}^{2}\approx \frac{d^{2}(\Omega _{1}^{2}-\Omega _{0}^{2})p}{%
\Omega _{0}^{2}+d^{2}-\Omega _{1}^{2}}.  \label{linear_p_width}
\end{equation}
When $\Omega _{1}$, however, happens to be close to $\Omega _{L}$, a
crossover is possible from the linear dependence (\ref{linear_p_width}) to
the square root behavior: 
\begin{equation}
\Delta _{im}^{2}\approx \sqrt{d^{2}(\Omega _{1}^{2}-\Omega _{0}^{2})p}.
\label{squareroot_p_width}
\end{equation}
The condition for such crossover to occur is 
\[
1\ll p\ll \frac{(\Omega _{0}^{2}+d^{2}-\Omega _{1}^{2})^{2}}{4d^{2}(\Omega
_{1}^{2}-\Omega _{0}^{2})}. 
\]

\subsubsection{The Lyapunov exponent and the density of states far from the
spectrum boundaries}

Using Eqs. (\ref{Re Lyamda}), (\ref{Im Lyamda}) and (\ref{FinalLyapun}) one
can calculate LE and the integral DOS for different regions of the spectrum.
For allowed bands, $(0,\Omega _{0})$, $(\omega _{il},\Omega _{1})$, $(\omega
_{pu},\infty ),$ one has 
\begin{eqnarray}
\lambda (\omega ) &=&-\frac{p(1-p)f^{2}}{8\left( pf-\gamma \right) }\frac{%
\omega }{c}\epsilon +O(\epsilon ^{2}),  \label{a-lamda} \\
N(\omega ) &=&\frac{1}{\pi }\frac{\omega }{c}\sqrt{\left| pf-\gamma \right| }%
+O(\epsilon ^{2});  \nonumber
\end{eqnarray}
for forbidden bands, $(\Omega _{0},\omega _{il}),(\Omega _{1},\omega _{pu}),$
we obtain 
\begin{eqnarray}
\lambda (\omega ) &=&\frac{\omega }{c}\left\{ \sqrt{pf-\gamma }-\left[ \frac{%
p(1-p)f^{2}}{8\left( pf-\gamma \right) }\right] \epsilon \right\} ,
\label{f-lamda} \\
N(\omega ) &=&0.  \nonumber
\end{eqnarray}
One can see from Eqs. (\ref{a-lamda}) and (\ref{f-lamda}) that within
allowed bands, DOS appears in the zero order of the formal expansion
parameter $\epsilon $, while LE starts from the first order. For the
forbidden bands the situation is reversed: LE contains a term of zero order
in $\epsilon $, while DOS in this order disappears. This observation
suggests a simple physical interpretation of the microcanonical
approximation in conjunction with the expansion over $\epsilon $. Analysis
of the series in this parameter shows that the actual small parameter is $%
a/(pl_{0})=l_{def}/l_{0}$, where $l_{def}$ is a mean distance between
impurities. The zero order expansion in this parameter can be interpreted as
a uniform continuous ($l_{def}\rightarrow 0$) distribution of impurities
with concentration $p$. The results in this approximation then corresponds
to the polariton impurity band which would exist in such a uniform system.
The parameter $l_{def}/l_{0}$ in this case is a measure of disorder in the
distribution of impurities, which leads to the localization of excitations
described by the DOS in Eq. (\ref{a-lamda}) with the localization length, $%
l^{-1}=$ $\lambda ,$ presented in the first line of the same equation.

Let us now consider frequencies in the interval $\omega \in \left( 0,\Omega
_{0}\right) \cup \left( \Omega _{L}^{i},\infty \right) $, where pass bands
of the pure chains composed of either host or impurity atoms overlap. The
DOS in this region can be written in a physically transparent form 
\begin{equation}
N(\omega )=\sqrt{\left( 1-p\right) N_{0}^{2}(\omega )+pN_{1}^{2}(\omega )}; 
\nonumber
\end{equation}
where \ $N_{0}(\omega )$ and $N_{1}(\omega )$ are integral DOS in pure
chains containing host atoms or impurities, respectively. In the remaining
portion of the initial spectrum $\omega \in \left( \omega _{up},\Omega
_{L}^{i}\right) $ (but not very close to the boundary $\omega _{up}$, where
the expansion ceases to be valid) DOS can be presented as $N(\omega )=\sqrt{%
\left( 1-p\right) N_{0}^{2}(\omega )-pl_{1}^{2}(\omega )}$, where $l_{1}$ is
the penetration length through the polariton gap of the 100\% impure chain.

Our main goal, however, is obtaining DOS and LE of the impurity induced
band, $\omega \in (\omega _{il},\Omega _{1})$. Near the center of this
region, the DOS can be presented in the following form 
\begin{equation}
N\left( \omega \right) =\frac{1}{\pi l_{0}}\left[ 1+\frac{4|\gamma |\omega
_{c}}{pd^{2}}\left( \omega -\omega _{c}\right) +O\left( \left( \frac{\omega
-\omega _{c}}{d}\right) ^{2}\right) \right] ,  \label{DOS in the center}
\end{equation}
where $\omega _{c}$ is the center of the impurity band, which in the linear
in $p$ approximation is 
\begin{equation}
\omega _{c}^{2}=\Omega _{1}^{2}-\frac{1}{2}\Delta _{im}^{2},
\label{omega in the center}
\end{equation}
where $\Delta _{im}^{2}$ is the width of the band in terms of squared
frequencies defined in Eq. (\ref{linear_p_width}). The first term in Eq. (%
\ref{DOS in the center}) represents the total number of states between $%
\omega _{il}$ and the center of the band; it is interesting to note that
this number does not depend upon concentration of impurities. The
coefficient at the second term gives the differential DOS at the center and
can be rewritten as 
\begin{equation}
\rho (\omega _{c})=\frac{4\left| \gamma \right| \omega _{c}}{\pi l_{0}pd^{2}}%
=\frac{2}{\pi l_{0}\delta },  \label{DDOSinthe center}
\end{equation}
where $\delta =\Delta _{im}^{2}/2\omega _{c}$ is an approximate expression
for the impurity band width $\delta \simeq \Omega _{1}-\omega _{il}$. This
DOS has a simple meaning of the average density of states uniformly
distributed through the entire band over the distance equal to $l_{0}/2$. In
one dimensional uniform systems the wave number of the respective
excitations, $k,$ is simply connected to $N\left( \omega \right) $: $k=$ $%
\pi N\left( \omega \right) $. The differential DOS would in this case be
proportional to the inverse group velocity. Accordingly, $1/\pi \rho (\omega
_{c})$ given by Eq. (\ref{DDOSinthe center}) can also be viewed as a group
velocity, $v,$ of excitations in the center of our impurity band in the case
of the uniform distribution of impurities. The expression for $v$ can also
be presented as 
\begin{equation}
v=\frac{d^{2}}{4\omega _{c}^{2}\left| \gamma \right| ^{3/2}}pc\ll c,
\label{group velocity}
\end{equation}
which demonstrates that polariton excitations of the impurity band have much
slower velocities not only compared to $c$ but also to the velocities at
both regular polariton branches.

Expanding Eq. (\ref{a-lamda}) for LE about the center of the band, we obtain
a parabolic frequency dependence of the localization length of the impurity
polaritons 
\begin{equation}
l\left( \omega \right) =\lambda \left( \omega \right) ^{-1}=2l_{0}\left(
pl_{0}\right) \left[ 1-\frac{20\gamma ^{2}}{p^{2}}\left( \frac{\omega
^{2}-\omega _{c}^{2}}{d^{2}}\right) ^{2}\right] .
\label{LL in the center in eps}
\end{equation}
One can see from this expression that $l\left( \omega \right) $ reaches its
maximum value $2pl_{0}^{2}$ at the center of the band. It is important to
note that the localization length here grows linearly with the
concentration, whereas it is LE that grows with concentration for
frequencies outside of the impurity band. An increase of the concentration
of the impurities also results in fast ($\propto 1/p^{2}$) flattening of the
maximum in the localization length, the fact we first noticed in our
numerical simulations.\cite{Blah}

The frequency dependence of LE, defined by Eqs. (\ref{Lyapunov}) and (\ref{y}%
), is shown in Fig. 1. For frequencies corresponding to the impurity band,
LE drops sharply (the localization length increases); it then diverges at
the upper boundary of the impurity band. This divergence has the same origin
as LE\ divergence at the lower boundary of the band gap of the original
crystal, where due to the specificity of the polariton spectrum, the wave
vector becomes infinite. The concentration dependence of LE and integral DOS
for some frequency $\omega _{0}\in (\Omega _{0},\Omega _{1})$ is shown in
Fig. 2. For $p=0$ this frequency belongs to the forbidden gap, thus DOS is
zero. With an increase of the concentration LE\ decreases (the localization
length increases). For the concentration when the lower boundary of the
impurity band crosses $\omega _{0},$ DOS becomes non-zero and in $\lambda
(\omega )$ the crossover between behaviors described by Eqs. (\ref{f-lamda})
and (\ref{a-lamda}) occurs. In Fig. 3 we compare the results of our
analytical calculations with numerical simulations of Ref. \onlinecite{Blah}%
. The comparison shows an excellent agreement between numerical and
analytical results, confirming the validity of the microcanonical method in
the considered limit. In addition, since numerical results were obtained
with the spatial dispersion taken into account, the comparison shows that
the model with dispersionless phonons produces reliable results for LE even
not very far away from the spectrum boundaries.

\subsubsection{The solution in the vicinity of the spectrum boundary.
Nonanalytical behavior.}

The results obtained in the previous section are clearly not valid for
frequencies close to the band boundaries $\omega _{il}$ and $\omega _{pu},$
where $pf-\gamma =0$ and the second term in the expansion (\ref{y expansion}%
) diverges. It is well known that perturbation expansions in disordered
systems usually fail in the vicinity of boundaries of the initial spectrum
of the system unperturbed by disorder.\cite{LGP} The regions in the vicinity
of these special frequencies $\omega _{il}$ and $\omega _{pu}$ require
special consideration. Attempting to regularize our $\epsilon $-expansion,
we shall seek corrections to the zero order solution in the form 
\begin{equation}
y=p+B\epsilon ^{\alpha }  \label{y alpha}
\end{equation}
admitting a possibility of fractional values of $\alpha $, and, hence,
non-analytical behavior of the solution. We also introduce a new variable $%
\zeta $ which determines the proximity to either of two frequencies $\omega
_{il}$ or $\omega _{pu}$: 
\begin{equation}
pf-\gamma =\zeta \epsilon ^{\alpha ^{^{\prime }}}.  \label{x}
\end{equation}
Substituting these expressions into Eq. (\ref{y}) and equating the lowest
order terms we see that the equation can only be satisfied for $\alpha
=\alpha ^{\prime }=2/3$. In this case we find that the parameter $B$
introduced in Eq. (\ref{y alpha}) obeys the equation 
\begin{equation}
\frac{p\left( 1-p\right) f}{2\sqrt{\zeta +Bf}}+B=0.  \label{B}
\end{equation}
The substitution of $y$ given by Eq. (\ref{y alpha}) in the condition for
the spectral boundaries presented by Eq. (\ref{y on a boundary}) allows one
to obtain an exact expression for the renormalized boundary 
\begin{equation}
pf-\gamma =\left[ 3p^{2/3}(1-p)^{2/3}\left( \frac{-f}{2}\right) ^{4/3}\right]
\epsilon ^{2/3},  \label{true boundary}
\end{equation}
that modifies Eq. (\ref{tentative boundaries}). The shift is small by virtue
of smallness of $\epsilon ^{2/3}$. In the lowest order in $\epsilon, $ the
new positions of the band edges for small $p$ are 
\begin{eqnarray}
\widetilde{\omega }_{il}^{2} &=&\Omega _{1}^{2}-\frac{d^{2}\left( \Omega
_{1}^{2}-\Omega _{0}^{2}\right) }{d^{2}-\left( \Omega _{1}^{2}-\Omega
_{0}^{2}\right) }p\left[ 1+3\left( \frac{a}{4pl_{0}}\right) ^{2/3}\right] ,
\label{renormalized boundary small p} \\
\widetilde{\omega }_{pu}^{2} &=&\Omega _{L}^{2}+\frac{d^{2}\left( \Omega
_{1}^{2}-\Omega _{0}^{2}\right) }{d^{2}-\left( \Omega _{1}^{2}-\Omega
_{0}^{2}\right) }p\left[ 1-3\left( \frac{a}{16pl_{0}}\right) ^{1/3}\right] ,
\nonumber
\end{eqnarray}
where we again encounter $a/(pl_{0})$ as a true small parameter of the
expansion. It is interesting to note the different character of
non-analytical corrections to the positions of the boundary of the impurity
band $\widetilde{\omega }_{il}^{2}$ and the bottom of the upper polariton
branch $\widetilde{\omega }_{pu}^{2}$. They have fractional concentration
dependence with the correction to $\widetilde{\omega }_{pu}^{2}$ being much
stronger in the limit $a/(pl_{0})\ll 1$.

Now let us consider the modification of LE and the integral DOS due to the
nonanalyticity. Substituting Eq. (\ref{y alpha}) into Eq. (\ref{Lyapunov})
the complex LE can be written as 
\begin{equation}
\widetilde{\lambda }=-\frac{p\left( 1-p\right) f}{2B}\epsilon ^{4/3}=\sqrt{%
\left( pf-\gamma \right) +Bf\epsilon ^{2/3}}\cdot \epsilon .
\label{Lyap nonanalitical}
\end{equation}
This expression explicitly demonstrates the crossover between analytical and
nonanalytiacal behavior. When frequency $\omega $ lies far from the initial
boundary $\left| pf-\gamma \right| \gg Bf\epsilon ^{2/3}$, one recovers the
term proportional to $\epsilon $ with the same coefficient as in Eq. (\ref
{FinalLyapun}), whereas in the opposite limit $\left| pf-\gamma \right| \ll
Af\epsilon ^{2/3}$, when we approach the boundary, the leading term gains
fractional power and becomes $\propto \epsilon ^{4/3}$.

At the vicinity of the renormalized spectral boundary, Eq. (\ref
{renormalized boundary small p}), DOS $\rho (\omega )\equiv 0$ to the left
of $\widetilde{\omega }_{il}$ and for $\omega >\widetilde{\omega }_{il}$ it
can be obtained as{\em \ } 
\begin{equation}
\rho \left( \omega \right) =\frac{3}{\pi d}\left( \frac{\gamma }{l_{0}}%
\right) ^{1/2}\frac{1}{(l_{0}p)^{1/2}}\frac{\omega }{\left( \omega ^{2}-%
\widetilde{\omega }_{il}^{2}\right) ^{1/2}},
\label{dos near rite eps-boundary}
\end{equation}
where $\gamma $ and $l_{0}$ are evaluated at $\omega =$ $\Omega _{1}$, and
we have neglected the renormalization of the boundary when calculating the
coefficient in $\rho \left( \omega \right) .$ In this approximation the
frequency and concentration dependence of DOS does not change as compared to
the one obtained from the non-renormalized expansion (\ref{a-lamda}). It is
interesting, however, that the renormalization brings about an additional
numerical factor of $3$ in Eq. (\ref{dos near rite eps-boundary}), which is
absent in the non-renormalized expression. This frequency dependence is
typical for excitations with the quadratic dispersion law in the long-wave
approximation. The main characteristics of this dispersion law is the
effective mass $m$, which can be found from Eq. (\ref{dos near rite
eps-boundary}) as 
\begin{equation}
m=\frac{9\gamma }{2pd^{2}l_{0}^{2}}=\frac{9\gamma ^{2}}{2pc^{2}}\left( \frac{%
\omega _{il}}{d}\right) ^{2}.
\end{equation}
It is interesting to compare this expression with the effective mass of the
upper polariton branch of the pure system at the bottom of the band $%
2m_{0}=(\omega _{L}/dc)^{2}$. The two expressions have a similar structure
if one introduces a ``renormalized speed of light'' 
\begin{equation}
\tilde{c}=c\sqrt{p/9\gamma ^{2}.}  \label{c_renorm}
\end{equation}
Though the introduced parameter $\tilde{c}$ does not have a direct meaning
of the speed of the excitations, it shows again that the excitations in the
impurity polariton band are considerably slower than their regular
counterparts, with a similar dispersion law at the spectrum boundary as well
as at the center of the band. Unlike, however, the center-of-band situation (%
\ref{group velocity}) the renormalized velocity at the edge is proportional
to the square root of concentration.

LE in the vicinity of $\widetilde{\omega }_{il}$ is represented by different
expressions for frequencies below and above $\widetilde{\omega }_{il}$
respectively 
\begin{eqnarray}
\lambda &=&\frac{1}{2l_{0}}\left( \frac{1}{4pl_{0}}\right) ^{1/3}\left[ 1+%
\frac{12}{d}\frac{\left( \gamma l_{0}\right) ^{1/2}}{\left( 4l_{0}p\right)
^{1/6}}\left( \widetilde{\omega }_{il}^{2}-\omega ^{2}\right) ^{1/2}\right] ,
\label{Lyapunov on the lower impurity bandedge} \\
\lambda &=&\frac{1}{2l_{0}}\left( \frac{1}{4pl_{0}}\right) ^{1/3}\left[ 1-%
\frac{12}{d^{2}}\frac{\gamma l_{0}}{\left( 4l_{0}p\right) ^{1/3}}\left(
\omega ^{2}-\widetilde{\omega }_{il}^{2}\right) \right] ,  \nonumber
\end{eqnarray}
reflecting a discontinuity of its frequency derivative at the spectrum
boundary. LE itself is, of course, continuous at $\widetilde{\omega }_{il}$,
giving rise to the localization length $l=2l_{0}\left( 4pl_{0}\right) ^{1/3}$
at the edge of the band. It is interesting to compare this expression with
the localization length at the center of the impurity band, Eq. (\ref{LL in
the center in eps}). They both grow with the concentration but the latter
one is much smaller and demonstrates slower fractional concentration
dependence.

At the upper impurity band edge $\Omega _{1}$ integral DOS, $N(\omega ),$
diverges causing much stronger singularity in the differential DOS than at
the lower boundary $\omega _{il}$ 
\begin{equation}
\rho \left( \omega \right) =\frac{d\omega p^{1/2}}{\pi c\left( \Omega
_{1}^{2}-\omega ^{2}\right) ^{3/2}},  \label{dos near left eps-boundary}
\end{equation}
which is typical for DOS in the vicinity of resonance frequencies. Comparing
this expression with a similar formula for a pure chain, one can again
interpret this result as a renormalization of the velocity parameter $c$ by
the concentration of impurities, which is different from that presented by
Eq. (\ref{c_renorm}) by a numerical coefficient only.

The last spectrum boundary is the bottom of the upper polariton branch $%
\Omega _{L}$. The spectrum in the vicinity of this frequency exists in the
absence of the impurities, which are responsible for two effects in this
region. First they move the boundary from $\Omega _{L}$ to higher
frequencies [Eq. (\ref{renormalized boundary small p})], and second, they
increase the effective mass of the upper polariton branch, such that the
differential DOS in the frequency region above $\widetilde{\omega }_{up}^{2}$
becomes 
\begin{eqnarray}
\rho \left( \omega \right) &=&\frac{1}{\pi dc}\frac{\omega }{\left( \omega
^{2}-\widetilde{\omega }_{up}^{2}\left( p\right) \right) ^{1/2}}
\nonumber \\
&\times & \left\{ 1+\frac{p}{2}\left[ \frac{d^{2}\left( \Omega _{1}^{2}-\Omega _{0}^{2}\right) }
{d^{2}-\left( \Omega _{1}^{2}-\Omega _{0}^{2}\right) }\right] ^{2}\right\} .
\label{dos on omege_up}
\end{eqnarray}
This expression can also be interpreted as a renormalization of the speed of
light $c.$

\section{Conclusion}

In the paper we have presented a detailed study of an impurity induced band
of excitations, which arise in the gap between lower and upper polariton
branches of a linear chain with dipole active atoms. We have also studied
impurity-induced effects on properties of the regular polariton branches.
The method of microcanonical ensemble in conjunction with the expansion in
the parameter $l_{def}/l_{0}\ll 1$, where $l_{def}$ is the average distance
between the impurities and $l_{0}$ is the localization radius of a single
local polariton state, produces a clear physical description of the
excitations, and shows excellent agreement with the results of numerical
simulations. In the zero order of this expansion, we recover DOS and the
dispersion law of excitations in the system with uniform continuously
distributed impurities. Corrections to this solution describe effects due to
local fluctuations in the positions of impurities such as a finite
localization length of the excitations, a renormalization of the spectral
boundaries and the effective mass of the excitations. The parameter $%
l_{def}/l_{0},$ therefore can be considered as a measure of disorder in the
system.

The main result of the calculations is the demonstration of the existence of
the polariton band formed by impurities without inner optical activity.
Justification for naming these excitations as polaritons lies in the fact
that they carry an electromagnetic field coupled with mechanical excitations
of the chain. The dispersion law of the excitations at the lower frequency
spectral boundary resembles that of the upper regular polariton band, but
with a significantly (by a factor of $p^{1/2}\ll 1$) reduced effective mass.
At the higher frequency edge of the band, the wave number diverges in a
manner similar to the regular lower polariton branch with a velocity
parameter again reduced by the same factor. The group velocity of the
excitations near the center of the band, as well as is the localization
length, is proportional to $p$. The latter, however, demonstrates a
parabolic frequency dependence with the curvature of the parabola falling
off with increase of the concentration as $1/p^{2}$. Excitations considered
in this paper are drastically different from impurity-induced polaritons
studied in Ref. \onlinecite{Singh}. The authors of the latter paper
considered excitations of an ordered chain of two-level atoms embedded in a
polar 1-d crystal. Most significantly, the effective mass of polaritons of
Ref. \onlinecite{Singh} is negative at the long-wave boundary of the
spectrum while our excitations have a positive effective mass in this
region; concentration dependencies of the effective mass and the band width
also differ significantly.

The regular expansion in powers of the parameter $l_{def}/l_{0}$ produces
diverging expressions at the boundaries of the zero order spectrum. Allowing
for fractional powers in the expansion, we obtained a finite regularized
expression for the density of states and the localization length at the
boundaries. It is interesting to note that renormalized DOS at the lower
boundary of the impurity band differs from its initial form only by a
numerical factor of $3$, while the position of the boundary is shifted
toward lower frequencies.

Experimental significance of the considered excitations is affected by two
factors: absorption due to different kind of anharmonic processes, and the
one-dimensional nature of the considered model. It is generally accepted
(see for instance Ref. \onlinecite{Lifshitz Kirpichnikov}) that one-dimensional
models give a fair description of tunneling in the limit of small
concentrations ($l_{def}/l_{0}\gg 1$). We deal with the opposite limit;
however, our zero order results could describe the dispersion law of
excitations in the real three dimensional medium propagating in one
specified direction (for example, the direction with the highest symmetry).
Disorder in this case would lead, of course, to scattering and deviation
from one-dimensional geometry, but effects due to disorder are much weaker
in three dimensions, and would not probably inhibit an observation of a
defect induced transparency in the frequency region of a polariton gap.

One can give a simple estimate for the role of absorption comparing the
absorption coefficient with the width of the impurity induced band. If the
latter is greater the band would manifest itself in transmission (or
reflection) experiments. Using Eq. (\ref{linear_p_width}) one can obtain the
following estimate for the admissible absorption coefficient $\alpha $: 
\[
\alpha \ll \frac{d^{2}(\Omega _{1}^{2}-\Omega _{0}^{2})p}{\Omega _{1}\left(
\Omega _{L}^{2}-\Omega _{1}^{2}\right) } 
\]
if the defect frequency $\Omega _{1}$ is not very close to the upper
boundary of the gap $\Omega _{L}$ or

\[
\alpha \ll \frac{d^{2}}{\Omega _{L}}\sqrt{p} 
\]
in the opposite case. The latter situation is much more favorable because of
the square root concentration dependence. For example, for frequencies of
the order of $1000$ $cm^{-1}$ and volume concentration of $0.1\%,$ which
correspond to linear concentration entering our formulas $p\sim $ $10\%$,
one has from the last expression $\alpha \ll 300$ $cm^{-1}$, while the
typical value of absorption for, e.g., alkali halides is $\sim 100$ $cm^{-1}$%
. This estimate shows that an observation of the transmittance through
polariton gaps due to impurity polaritons is feasible.

\section{Acknowledgment}

We are pleased to acknowledge valuable remarks of A.J. Sievers. We wish to
thank S. Schwarz for reading and commenting on the manuscript. This work was
supported by the NSF under grant No. DMR-9632789 and by a CUNY collaborative
grant.

\begin{figure}[htb]
\caption{LE for a chain with a defect concentration of 10 \% (solid
line) in comparison to a pure system (dashed line).}
\end{figure}

\begin{figure}[htb]
\caption{Dependences of LE (solid line) and DOS (dashed line) on
concentration for a frequency in the interval $\Omega_0 < \omega < \Omega_1$.}
\end{figure}

\begin{figure}[htb]
\caption{Comparison of LE calculated with (dashed line) and without
(solid line) spatial dispersion in the vicinity of $\Omega_1$. The
concentration of defects for both curves is 1 \%.}
\end{figure}

\end{multicols}
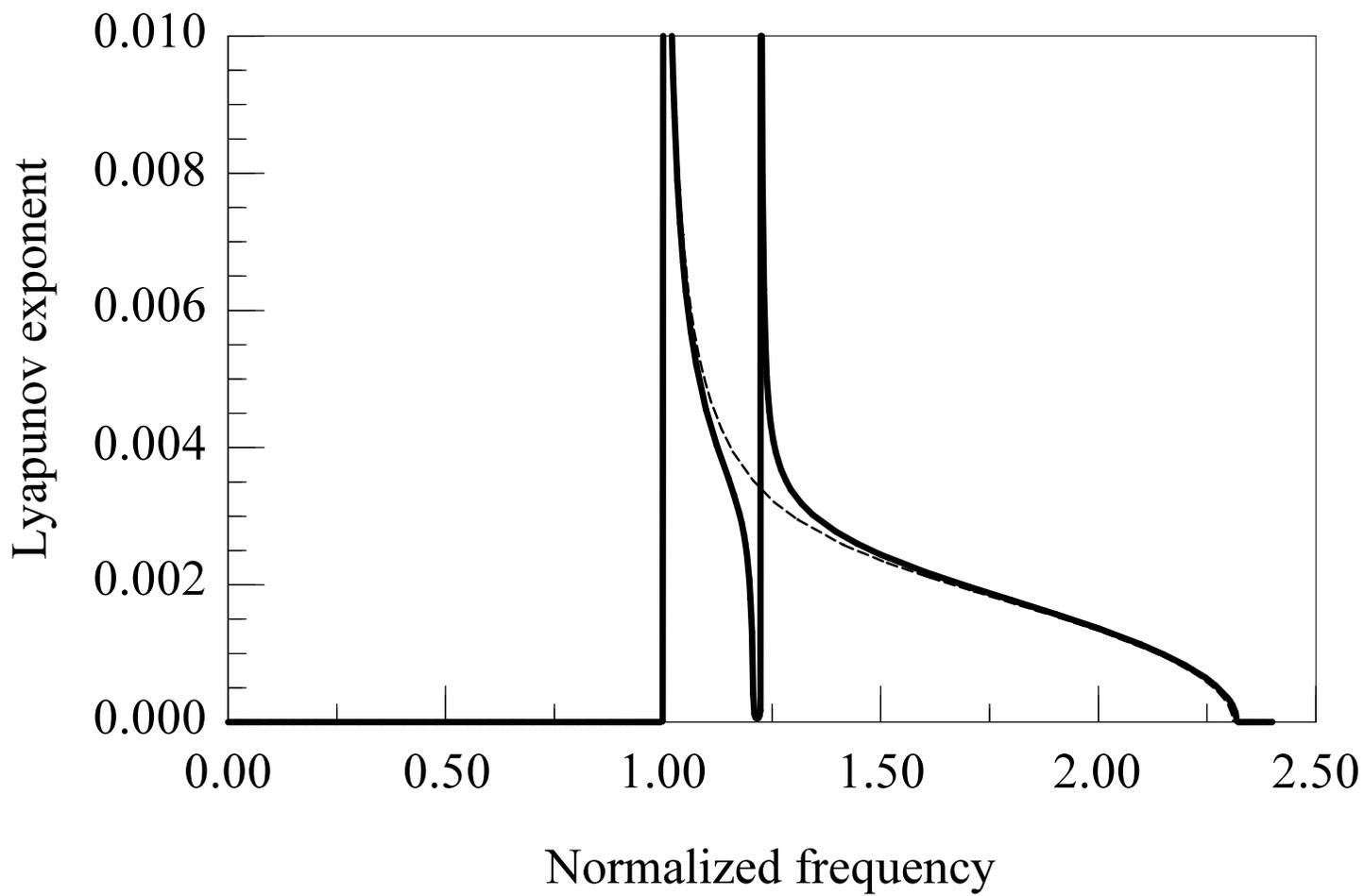
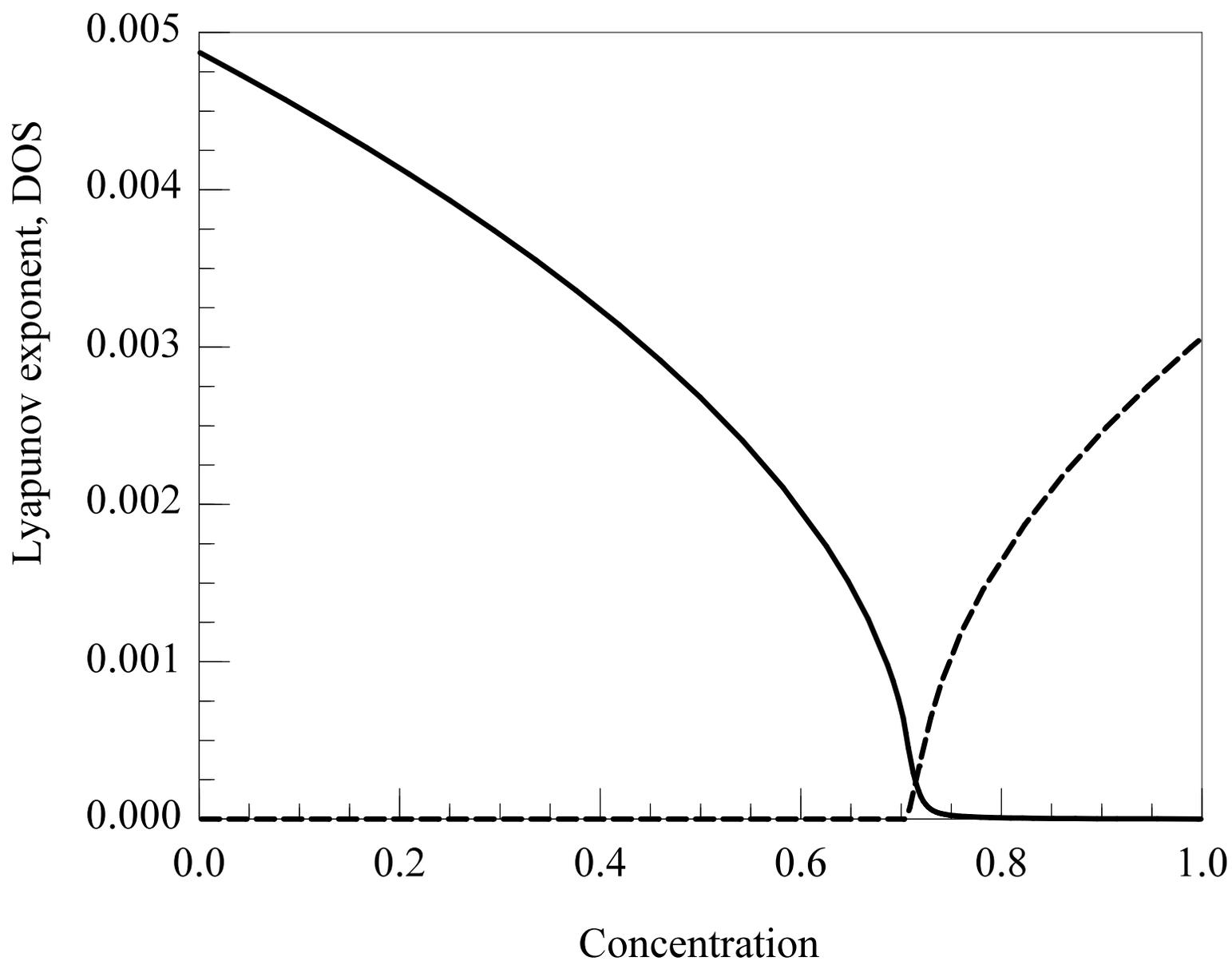
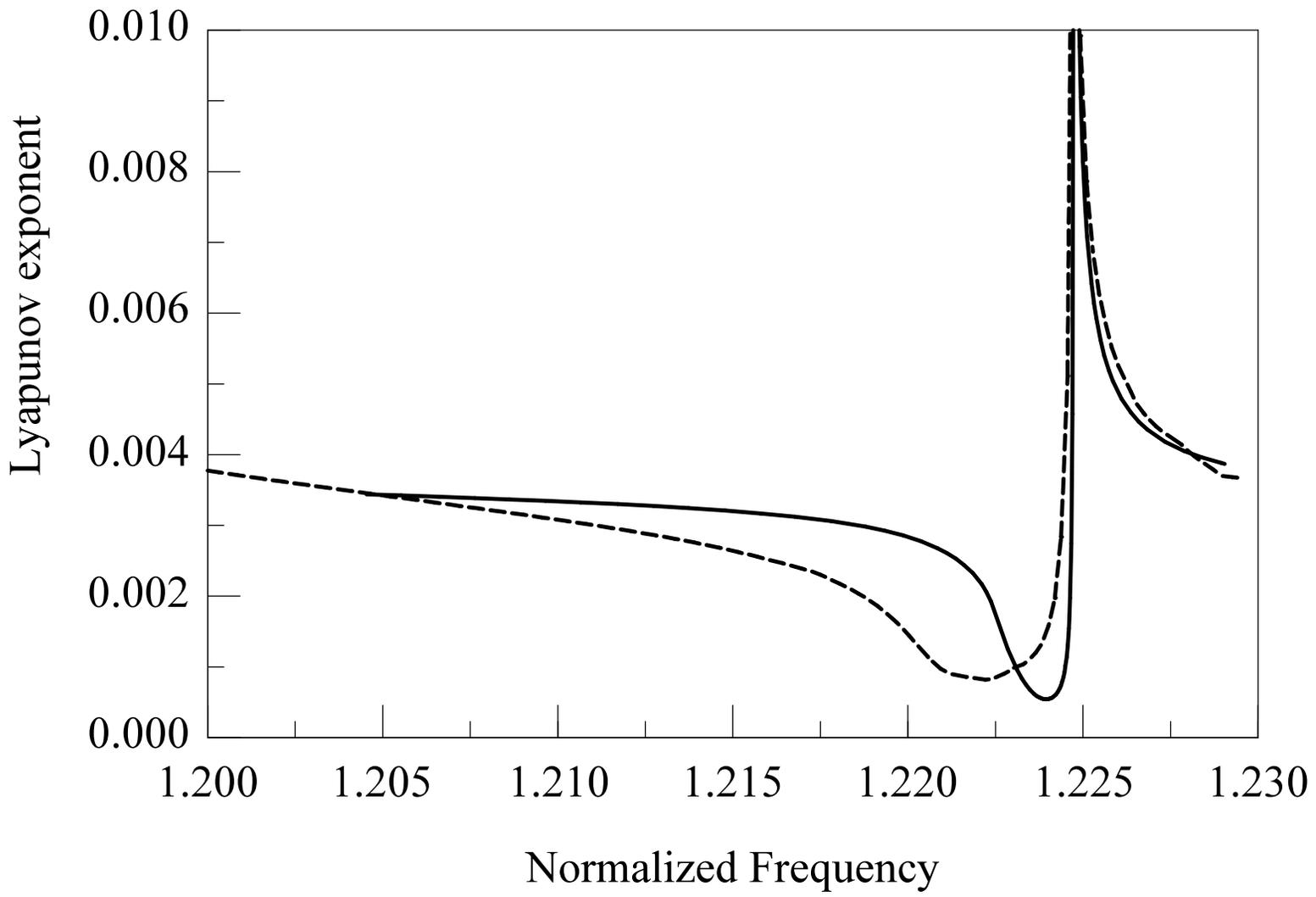
\end{document}